\documentclass[10pt,journal,compsoc]{IEEEtran}

\ifCLASSOPTIONcompsoc
  \usepackage[nocompress]{cite}
\else
  \usepackage{cite}
\fi

\ifCLASSINFOpdf
\else
\fi

\usepackage{url}
\usepackage{hyperref}
\usepackage[utf8]{inputenc}
\usepackage{makecell}
\usepackage{multirow}
\usepackage{verbatim}
\usepackage{xspace}
\usepackage{xcolor}

\usepackage{graphicx}
\usepackage{pdflscape}
\usepackage{array}

\usepackage[breakable, theorems, skins]{tcolorbox}
\usepackage{balance}
\usepackage{fontawesome}
\usepackage{adjustbox}
\usepackage{rotating,tabularx}
\usepackage{subfiles}

\definecolor{darkgreen}{rgb}{0.05,0.5,0.05}

\newcommand{\ghtorrent}{GHTorrent\xspace}
\newcommand{\gh}{GitHub\xspace}

\tcbset{enhanced}

\DeclareRobustCommand{\mybox}[2][gray!13]{%
\begin{tcolorbox}[   %
        breakable,
        left=1pt,
        right=1pt,
        top=1pt,
        bottom=1pt,
        colback=#1,
        colframe=black,
        width=\dimexpr\columnwidth\relax, 
        enlarge left by=0mm,
        boxsep=5pt,
        ]
        #2
\end{tcolorbox}
}

\renewenvironment{quote}{%
   \list{}{%
     \leftmargin\parindent%
     \rightmargin0cm
   }
   \item\relax
}
{\endlist}

\newcommand{\inlinequote}[2]{\emph{``#1''} (P{#2})\xspace}
\newcommand{\blockquote}[2]{\begin{quote}{\small\faComment}~\emph{#1} (P{#2})\end{quote}}

\begin{document}
\title{Including Everyone, Everywhere:\\ Understanding Opportunities and Challenges of Geographic Gender-Inclusion in OSS}

\author{Gede Artha Azriadi Prana, Denae Ford, Ayushi Rastogi,\\ David Lo, Rahul Purandare, Nachiappan Nagappan}

\markboth{IEEE Transactions on Software Engineering, 2021}%
{Shell \MakeLowercase{\textit{et al.}}: Bare Demo of IEEEtran.cls for Computer Society Journals}

\IEEEtitleabstractindextext{%
\begin{abstract}
The gender gap is a significant concern facing the software industry as the development becomes more geographically distributed.
Widely shared reports indicate that gender differences may be specific to each region. However, how complete can these reports be with little to no research reflective of the Open Source Software (OSS) process and communities software is now commonly developed in?
Our study presents a multi-region geographical analysis of gender inclusion on GitHub. This mixed-methods approach includes quantitatively investigating differences in gender inclusion in projects across geographic regions and investigate these trends over time using data from contributions to 21,456 project repositories.
We also qualitatively understand the unique experiences of developers contributing to these projects through a survey that is strategically targeted to developers in various regions worldwide.
Our findings indicate that gender diversity is low across all parts of the world, with no substantial difference across regions. However, there has been statistically significant improvement in diversity worldwide since 2014, with certain regions such as Africa improving at faster pace. We also find that most motivations and barriers to contributions (e.g. lack of resources to contribute and poor working environment) were shared across regions, however, some insightful differences, such as how to make projects more inclusive, did arise.
From these findings, we derive and present implications for tools that can foster inclusion in open source software communities and empower contributions from everyone, everywhere.

\end{abstract}

}

\maketitle

\IEEEdisplaynontitleabstractindextext

\IEEEpeerreviewmaketitle

\IEEEraisesectionheading{\section{Introduction}\label{sec:introduction}}

The gender gap in the software industry is alarming, garnering attention worldwide.
IT companies in India reportedly have women concentration in lower career levels~\cite{raghuram2017women}.
In the United States, women earning computing degrees rose since the mid-1990s, yet they comprise a quarter of computing professionals~\cite{dubow2018ncwit}.
An estimate by the European Commission~\cite{ec-increaseingendergap} suggests that if more women enter the digital job market, it could create an annual EUR 16 billion GDP boost for the European economy. 

Similar investigations in open source software systems show that despite no significant differences between the work practices of men and women~\cite{canedo2020work} and improved team performance in gender-diverse teams~\cite{ortu2017diverse}, women make up less than 10\% of core contributors~\cite{bosu2019diversity}. 
Further, horizontal and vertical segregation exist~\cite{canedo2020work}. 

In open source, explorations on gender diversity are all-inclusive, implicitly assuming that the problem remains the same irrespective of the population and project characteristics.
However, in this approach, we are likely to miss local achievements in promoting gender diversity and/or problems unique to others. 
One factor to consider is the geographical region. 
A study conducted within the European Union shows a disparity in women's participation in digital economies, with Finland and Sweden scoring the highest while Greece and Italy the lowest~\cite{ec-increaseingendergap}. 
This example suggests that digital and online engagement can shift across geographic regions in addition to genders. Thus, inspiring us to ask how this difference in engagement can manifest in open source, specifically.

Our study presents the largest exploration into gender diversity in open source software projects in different parts of the world.
We investigate active and collaboratively developed software projects hosted on GitHub to answer:

\emph{RQ1: What are the gender and geographic diversity characteristics of open source software projects on GitHub?}

The first question is exploratory, presenting the state-of-the-practice on gender diversity and substantiating the need for exploration.
Further, we ask questions to open source software contributors to understand:

\emph{RQ2: What factors potentially contribute to the differences in gender and geographic-based developer participation?}

Our analysis is based on 21,456 carefully selected software projects on GitHub.
We use a sequential mixed-methods approach.
First, we quantitatively analyze archived software engineering data of the selected projects to show the state-of-practice of gender diversity worldwide.
Next, we survey 1,562 contributors, strategically identified from the selected projects based on gender and geography.
We solicit their response in search of factors that can potentially contribute to the differences in developer participation based on gender and geography worldwide.

Our analyses of a decade of development activities on GitHub show small but significant improvements in gender diversity in the last five years.
While we celebrate the positive change, it is important to remember that we are far from reaching gender balance.
Our study further shows that gender diversity changes over time have not been the same across regions.
Some regions such as Eastern Asia and Northern America are (relatively) ahead in gender diversity, while others such as Eastern Europe and Sub-Saharan Africa are still catching up.
These differences are also reflected in our investigation of gender and regional related motivations and challenges.

This comprehensive guide of gender-geographic challenges and opportunities can direct future in-depth explorations catering to sub-population needs. 
For example, one of the opportunity identified here is having a code of conduct. 
Having a code of conduct can support a two-pronged approach of: 1) allowing lurkers interested in contributing (e.g., including women and other marginalized developers) to feel more comfortable in contributing since they know there are guidelines that can protect them from toxic interactions and 2) signal to developers who are already in the community (e.g., including those that may have been inciting toxic interactions) that there will be repercussions for their actions.
Solutions such as these can have a long-term impact to minimize gender gap and uplifting society.

Our contributions are as follows:
\begin{enumerate}
    \item We present an analysis of the activity and experiences at the intersection of gender and global geographic region.
    \item Large-scale global analysis of regional gender diversity spanning 21,456 active GitHub repositories and 70,621 commit authors.
    \item Global survey of factors that contribute to the differences in gender and geographic-based developer participation, with 122 respondents across 5 large geographic regions and across genders.
    \item A discussion of actionable implications of how to support OSS sub-communities across gender and geographic regions.
    \item A publicly available dataset to encourage further investigations.
\end{enumerate}

\section{Background and Related Work}\label{sec:background}

Success of open source software projects is attributed to its developers.
This inspired a series of studies exploring reasons for open source engagement.
These studies include motivations for developer participation~\cite{roberts2006understanding}, barriers to participation~\cite{steinmacher2015social}, and how developers contribute to open source~\cite{zlotnick2017github}. 
These studies help understand and optimize the opportunities to retain community participation.
It also prepares projects to avoid or mitigate situations that causes contributors to leave projects. 

This paper is inspired by and extends works on motivation and barriers to participation in open source software projects along the lines of diversity in terms of gender and region of contributors in software projects. 
Next, we present important studies that have shaped this area of research.

\subsection{Motivation to contribute to open source software.}
Motivation in software engineering has been subject to numerous studies, including several systematic reviews~\cite{beecham2008motivation,francca2011motivation,francca2018motivation}. The existing body of works include a number of studies focusing specifically on motivation of OSS contributors. For example, in a 2002 study, Hars and Ou~\cite{alexander2002working} surveyed open source developers and found that their motivation for contributing are diverse -- while students and hobbyists tend to be internally motivated, there are also a large number of developers who are motivated by external rewards. Lakhani and Wolf~\cite{lakhani2003hackers} surveyed 684 OSS developers and found that the strongest type of motivation among the respondents are enjoyment-based intrinsic motivation. prominence of enjoyment-based intrinsic motivation. Von Krogh et al.~\cite{von2012carrots} examined prior literature on OSS developers' motivation to contribute, and proposed 10 clusters of motivation types categorized into intrinsic motivation, internalized extrinsic motivation, and extrinsic motivation. Barcomb et al.\cite{barcomb2019episodic} surveyed episodic (non-habitual) OSS volunteers and found that intention to remain are positively associated with social norms, satisfaction, and community commitment. Further, they also found some differences based on participants' gender. Most recently, a study by Gerosa et al.~\cite{gerosa2021shifting} investigated how main motivations of OSS contributors as a group change over the years and how OSS contributors' individual motivations change as they become more experienced. They found that among OSS contributors, some motivations related to social aspect has gained popularity in recent years. They also found that experienced OSS contributors tend to be motivated by intrinsic factors such as altruism, unlike new contributors who tend to place higher importance on factors such as career and learning. These studies facilitate better understanding of what drives people to contribute to OSS projects, what approaches project owners can take to attract contributors, and how these contributors can be retained.

\subsection{Barriers to participation in open source.}
A number of studies investigate barriers that can prevent developers from participating to open source.
These barriers have been identified in tools, processes~\cite{mendez2018open}, and social collaborations~\cite{steinmacher2015social}. For example, a study by Terrell et al.~\cite{terrell2017gender} found that while women have higher overall acceptance rate of pull requests, their acceptance rate is lower than men when their gender are identifiable and they are not insiders to a project. Another study by Rastogi et al.~\cite{rastogi2018relationship}, which analyzes pull requests from 17 countries, found that acceptance rate of contributions can vary significantly depending on the contributor's country of origin, and are higher when when they are evaluated by developers from the same country. The  study  however  does  not  analyze  gender as a factor, as they noted that including only pull requests for which gender data can be obtained will result in sample size that is too small. Other studies examine barriers such as those affecting acceptance of contribution from newcomers~\cite{steinmacher2015social} or those affecting underrepresented communities~\cite{ford2019remote}). 
These studies not only help in raising awareness of existence of such barriers, but they also help in identifying the source of problem. Further, studies such as~\cite{steinmacher2016overcoming} also propose solutions that can be adopted by OSS community to mitigate such barriers.

\subsection{Diversity in open source software projects.} 
In line with increasing awareness regarding the importance of diversity in broader work context, diversity in open source software projects has gained increasingly widespread attention.
Starting from the awareness of diversity and particularly the demographic attributes of developers~\cite{vasilescu2015perceptions, robles2016women}, today improving diversity is seen as a goal for fairness~\cite{terrell2017gender} as well as improved productivity~\cite{vasilescu2015gender}.
Many studies relating to gender diversity and the lack thereof followed, discussing its relevance~\cite{terrell2017gender}, state of diversity among popular OSS projects~\cite{bosu2019diversity}, male and female OSS contributors' perceptions of other contributors~\cite{lee2019floss}, perceptions of women core developers in OSS projects~\cite{canedo2020work}, and the impediments to improve gender diversity~\cite{imtiaz2019investigating}.

All these studies identify challenges and needs of under-represented communities.  We conduct a comparison outlining the distinction between our work and closely-related prior work in %
Table~\ref{tab:relatedwork} in Appendix~\ref{sec:lit-table}.

Our study has common elements to the developer survey on Stack Overflow users~\cite{zlotnick2017github} but their findings are not synonym to open source. 
That said, the report does not provide empirical data to support the full scope of motivations and how they persist across genders or regions. 
Our work provides novelty by conducting an analysis of the activity and experiences at the intersection of gender and global geographic region.
Taking research on the subject a step further, in this work we study gender diversity in different regions and how factors relating to gender and region can potentially explain why developers join open source software projects, select a project, continue participation. Such factors can potentially also explain barriers and reasons to leave a project.

\section{Methodology} \label{sec:methodology}

We used a convergent mixed-methods study approach to answer our research questions~\cite{creswell2018research}. 
We identified active OSS projects that are likely to be non-toy projects, resolved the gender as well as location of the project contributors, and then distributed a survey to understand their motivations and challenges. The following subsections describe each of these steps in detail.

\subsection{Identification of Suitable \gh Repositories}
\subsubsection{Initial Set of Repositories}
We chose to use \ghtorrent data as it has been widely used in software engineering research, including in works related to diversity (e.g.~\cite{vasilescu2015gender, ortu2017diverse, terrell2017gender}). Using the latest \ghtorrent database dump (1 June 2019), we begin by filtering for repositories that are active, are not toy repositories, and involve collaboration between different developers. We use the following repository criteria:
\begin{itemize}
    \item The repository has existed for at least 180 days (measured using difference of \textit{updated\_at} and \textit{created\_at} columns in the \ghtorrent data). This is to reduce probability that the project is a ``toy'' repository (e.g., a user trying a programming tutorial) or a student programming assignment (which usually lasts less than a semester).
    \item The repository has at least one commit from the beginning of 2018 or later. This is to reduce probability that the project is inactive.
    \item The repository has at least 10 commits from 4 or more distinct commit authors, none of which are marked `fake' or `deleted' \ghtorrent.
    \item The repository is not a fork. We chose not to evaluate forks since we are interested in ``core'' contributors of a project. In addition, contributions to forks are not always integrated back to the original project and there may also be redundant development between forks and original projects~\cite{zhou2018identifying}.
\end{itemize}

The above criteria were set to reduce probability of including ``toy'' projects while avoiding potential elimination of active non-toy projects. Considering rapid growth of GitHub in recent years, we believe the criteria still allows newer OSS projects, for example those created in 2018, to be included in the study.

\subsubsection{Location Resolution of Commit Authors}
We subsequently attempt to resolve the location of the commit authors. As \ghtorrent data does not include personal information, we collect additional information through the GitHub API prior to location and gender resolution. For location, resolution is based on value of \textit{country\_code} field of the commit author's user information, if available. If the field is empty, location resolution is attempted using other fields in the following order:
\begin{enumerate}
    \item \textit{location} field. For example, if the commit author specifies ``Seattle'' as their location, the country assigned will be USA. If they specifies "Tokyo", the country assigned will be ``Japan''.
    \item Latitude and longitude (\textit{lat} and \textit{long} fields in \ghtorrent data, respectively).
    \item \textit{company} field. For example, ``Argonne National Lab'' or ``Puget Sound Regional Council'' are considered as evidence that the commit author is based in the USA. ``German National Library'' is considered as evidence that the author %
    is based in Germany. Where possible, we attempt to resolve an organization's location using its website and LinkedIn page. In case of multinational organizations, the author's location is considered unresolved unless more specific information such as branch name is provided. For example, ``RedHat'' will be considered as unresolved location, whereas ``RedHat UK'' will be considered as evidence that the %
    location is the UK.
    \item \textit{email} field. For example, if the author's email address uses an Australian government domain, the country assigned will be Australia.
\end{enumerate}

Considering differences in culture and other factors that may exist within a region (for example, North American countries versus Latin American countries, Western European countries versus Eastern European countries), we also assign three levels of region information to each commit author based on the taxonomy of regions specified by United Nations Statistics Division\footnote{https://unstats.un.org/unsd/methodology/m49/}. For example, if the commit author's resolved location is Kenya, the assigned region information will be ``Africa'' (region level 1), ``Sub-saharan Africa'' (region level 2), and ``Eastern Africa'' (region level 3). Our intention is to facilitate analyses at finer granularity instead of treating a continent (e.g. America, Asia, Europe) as a unit.

\subsubsection{Gender Resolution of Commit Authors}
For the commit authors' gender, resolution is attempted by identifying first name portion of the commit author's name. This is followed by resolution of gender using \textit{genderize.io}\footnote{http://www.genderize.io}, which has been reported to have high accuracy~\cite{santamaria2018comparison,karimi2016inferring} and has been used in various studies related to gender representation (e.g.~\cite{holman2018gender, thomas2019gender, russell2018large}) as well as in the media\footnote{https://genderize.io/use-cases} For this part, titles (e.g. ``Dr.'') are ignored, and if the commit author does not use Latin alphabet to specify their name, the name is first converted to Latin alphabet using a combination of CC-CEDICT\footnote{https://cc-cedict.org/wiki/} (for Chinese characters) and Google Translate\footnote{https://translate.google.com/}.

As an additional measure to evaluate \textit{genderize.io}'s accuracy, one of the authors randomly selected five sample repositories for manual validation. The repositories are associated with a total of 57 contributors from different regions (15 from Americas, 12 from Asia, 18 from Europe, 4 from Oceania, and 8 with unknown region). Each repository is assigned to each of the remaining authors who subsequently attempt manual gender resolution using public information sources (the contributor's GitHub page, LinkedIn page, Twitter profile, etc.). The result is subsequently compared to gender prediction result from \textit{genderize.io}. We find that overall the manual analysis results match \textit{genderize.io}'s results 89.5\% of the time, with 100\% match on European and Oceanian contributors, 91.7\% on Asian contributors, 80\% on contributors from Americas. In case of contributors whose location is unresolvable, there is 75\% agreement between manual resolution and \textit{genderize.io}'s prediction based on contributors' names.

\subsubsection{Final Selection of Repositories}
Following this, we apply further filtering for repositories for which both gender and location can be resolved for at least 75\% of the commit authors. Considering that not all repositories on GitHub are software project repositories~\cite{kalliamvakou2014promises}, we also exclude repositories for which GitHub detects no primary language.%
In all, after the entire process, 21,456 repositories are shortlisted, with the breakdown of filtering result at various stages shown in Table~\ref{tab:repository_filtering}. We also extract all commit authors associated with the shortlisted repositories. Tables~\ref{tab:overview_repo} and~\ref{tab:overview_developers} show the statistics of the dataset.

\begin{table}[!htb]
    \caption{Result of project repository filtering steps.}
    \label{tab:repository_filtering}
    \centering
    \begin{tabular}{|l|r|}
        \hline
        Filtering step & Count\\
        \hline
        Initial number of repositories & 125,485,095 \\
        \makecell[tl]{Repositories with commits\\newer than January 1, 2018} & 31,947,039\\
        \makecell[tl]{Repositories that have existed\\for at least 180 days\\and are not marked as ``deleted\\} & 4,393,507\\
        \makecell[tl]{Repositories with at least 10 commits,\\and are not a fork} & 2,129,448\\
        \makecell[tl]{Repositories remaining with no commit authors\\marked ``fake'' or ``deleted''} & 97,989\\
        \makecell[tl]{Repositories with 75\% commit authors\\having resolvable gender and location} & 21,456\\
        \hline
    \end{tabular}
\end{table}

\begin{table}[!htb]
    \caption{Statistics of shortlisted repositories and associated commit authors.}
    \label{tab:overview_repo}
    \centering
    \begin{tabular}{|l|r|r|r|r|}
    \hline
    \multicolumn{5}{|c|}{Shortlisted Repositories}\\
    \hline
    & \makecell[tc]{Min} & \makecell[tc]{Max} & \makecell[tc]{Mean} & \makecell[tc]{Median}\\
    \hline
    No. of Commit Authors & 4 & 109 & 6.16 & 5\\
    No. of Commits & 22 & 301692 & 363.27 & 170\\
    Creation year & 2008 & 2018 & 2014.63 & 2015\\
    \hline
    \multicolumn{5}{|c|}{Commit Authors of Shortlisted Repositories}\\
    \hline
    \multicolumn{4}{|l|}{Total commit authors count} & 70,621\\
    \multicolumn{4}{|l|}{Commit authors with resolvable location} & 58,498\\
    \multicolumn{4}{|l|}{Commit authors with resolvable gender} & 65,132\\
    \multicolumn{4}{|l|}{Commit authors with resolvable gender and location} & 56,866\\
    \hline
    \end{tabular}
\end{table}

\setlength\tabcolsep{3pt}
\begin{table}[!htb]
    \caption{Commit author region and gender in shortlisted repositories, sorted by Region Level 1.}
    \label{tab:overview_developers}
    \centering
\begin{tabular}{|l|l|r|r|r|r|}
\hline
\multirowcell{2}{Region\\Level 1} & \multirowcell{2}{Region\\Level 2} & \multirow{2}{*}{Count} & \multicolumn{3}{|c|}{Percentage}\\
\cline{4-6}
& & & Man & Woman & \makecell[tl]{Un-\\known}\\
\hline
 Africa & Northern Africa & 91 & 91.21 & 5.49 & 3.33 \\
 Africa & \makecell[tl]{Sub-Saharan Africa}  & 273 & 92.67 & 3.66 & 3.66 \\
 Americas & \makecell[tl]{Latin America and\\the Caribbean} & 2547 & 93.29 & 4.75 & 1.96 \\
 Americas & \makecell[tl]{Northern America} & 24055 & 90.27 & 7.47 & 2.25\\
 Americas & Others & 5 & 80.00 & 0.00 & 20.00 \\
 Asia & Central Asia & 34 & 88.24 & 2.94 & 8.82 \\
 Asia & Eastern Asia & 2585 & 80.46 & 10.10 & 9.44 \\ 
 Asia & \makecell[tl]{South-eastern Asia} & 686 & 87.90 & 6.85 & 5.25\\
 Asia & Southern Asia & 1463 & 91.46 & 5.47 & 3.08\\
 Asia & Western Asia & 529 & 93.19 & 3.40 & 3.40\\ 
 Europe & Eastern Europe & 3858 & 94.35 & 2.90 & 2.75\\ 
 Europe & Northern Europe & 7541 & 92.71 & 5.38 & 1.91\\ 
 Europe & Southern Europe & 2314 & 94.77 & 3.11 & 2.12\\
 Europe & Western Europe & 10637 & 92.94 & 3.88 & 3.18\\ 
 Oceania & \makecell[tl]{Australia and\\ New Zealand} & 1870 & 92.62 & 5.13 & 2.25\\ 
 Oceania & Melanesia & 5 & 80.00 & 0.00 & 20.00\\
 Oceania & Polynesia & 5 & 100.00 & 0.00 & 0.00\\
 Unknown & Unknown & 12123 & 61.96 & 6.22 & 31.82\\ 
 \hline
\end{tabular}
\end{table}
\setlength\tabcolsep{6pt}

\subsubsection{Calculating Gender Diversity of Commit Authors}
To measure the gender diversity of commit authors from different regions, we use the Blau diversity index~\cite{blau1977inequality} which has also been used in several works in software engineering  domain~\cite{vasilescu2015gender,vasilescu2015data,catolino2019gender}. In simple terms, the index specifies the probability that two randomly-selected members of a group would belong to different categories. It is defined as $1 - \sum_{i \in \{m,f\}} {p_i^2 }$, where $p_i^2$ are proportion of men and women (``M'' and ``F'', respectively) among commit authors.

During calculation, we disregard unknown values. For example, if a region is associated with five commit authors, and four of them are identified as men while one is unknown, the gender diversity index will be 0. Similarly, if a set of commit authors from a region comprise two men, two women, and one person with unidentified gender, the gender diversity index will be 0.5, which is the maximum value. 

To check whether the diversity of commit authors is independent from region, we apply the Chi-squared test to analyze distribution of the two genders across regions, and subsequently computed Cramér's V~\cite{cramer1946mathematical} to measure association strength between gender and region at both region levels. For this analysis, we include commit authors whose location and gender are resolvable (56,866 commit authors comprising 53,426 men and 3,440 women). Exclusion of commit authors with unknown gender is done for consistency with Blau diversity index computation, while exclusion of commit authors with unknown location is done since we are interested in variation between regions worldwide. 

Since we note that most projects (70.27\%) have a majority region at region level 1, i.e. level 1 region from which more than half commit authors originate, we also performed a repository-oriented diversity analysis to provide additional perspective. To do this, we first associate a repository to a location based on the most common identified location of the commit authors. For example, if five commit authors contribute to a repository, and their locations are \{``Europe'', ``Americas'', ``Americas'', ``Americas'', ``Asia''\}, then the repository will be associated with Americas. Afterwards, we compute the diversity index of each repository. To test statistical significance and effect size of the difference, we first apply Kruskal-Wallis H test on groups of repositories associated with each level 1 regions. We subsequently applied Mann-Whitney U test~\cite{mann1947test} with Bonferroni correction~\cite{abdi2007bonferroni} to compare different pairs of region level 1. Afterwards, we computed Cliff's Delta~\cite{cliff2014ordinal} on level 1 region pairs\footnote{We use https://github.com/neilernst/cliffsDelta implementation for Cliff’s Delta test} with statistically significant difference to discover the effect size. 

After we conducted our initial analysis on the commit authors, we also considered following the line of research of Trinkenreich et al.~\cite{trinkenreich2020hidden} by investigating activities of non-technical contributors. We extracted data of \gh users who had never authored a commit to the shortlisted sample repositories but had created, changed, or commented on issues and merged pull requests associated with the sample repositories. We excluded user IDs that are marked ``fake'' and ``deleted'' in \ghtorrent. We found 299,159 users that are not also commit authors. Out of this group, 30.59\% has both unresolvable gender and location. Beyond this, 21.56\% has unresolvable location although their genders are resolvable, and 9.54\% has unresolvable gender although their locations are resolvable. Table~\ref{tab:nonauthor_diversity_by_region_level_1} shows the breakdown of this non-author group by region level 1, along with the Blau index of the users in this group whose gender is resolvable. We note that for members of this group with resolvable gender and location, the vast majority is male, and like the case with commit authors, there is low diversity in the various regions studied. However, due to the large percentage of users with unknown gender and/or location among this group, we decided not to analyze this group and to focus our analysis solely on commit authors. 

\begin{table}[!htb]
    \caption{Diversity and counts of contributors other than commit authors by region level 1. Entries are ordered by non-decreasing Blau index value. Blau index of 0.5 indicate maximum diversity (50\% men, 50\% women)}
    \label{tab:nonauthor_diversity_by_region_level_1}
    \centering
\begin{tabular}{|l|r|r|r|r|r|r|r|}
\hline
\multirowcell{2}{Region\\Level 1} & \multicolumn{4}{|c|}{Count} & \multirowcell{2}{\%}& \multirowcell{2}{Blau\\index} \\
\cline{2-5}
& M & W & Unknown & Total & & \\
\hline
Europe& 43873 & 1402 & 10303 & 55578 & 18.6 & 0.06 \\
Oceania & 3359 & 121 & 1022 & 4502 & 1.5 & 0.07  \\
Americas& 44859 & 2430 & 8909 & 56198 & 18.8  & 0.10 \\
Africa & 1432 & 87 & 387 & 1906 & 0.6 & 0.11 \\
Asia & 15424 & 1351 & 7860 & 24635 & 8.3 & 0.15 \\
\hline
Unknown & 59486 & 4857 & 91428 & 155771 & 52.2 & 0.14 \\
\hline
\end{tabular}
\end{table}

\subsubsection{Examining Correlation between Geographic and Gender Diversity}
We are also interested in whether a repository's gender diversity correlates with its geographic diversity. As the Blau index values of repositories' contributor gender and location diversity are not normally distributed (D'Agostino's K$^2$ test~\cite{d1990suggestion} yields p=0.00 for gender diversity index values as well as region diversity index for all levels of regional grouping), we analyze this by computing Spearman's rank correlation test~\cite{hopkins1997new} between repositories' gender diversity index values and geographic diversity index values at different regional groupings. We use \textit{SciPy}~\cite{2020SciPy-NMeth} implementation of these statistical tests, and follow scale of interpretation of $\rho$ used by Camilo et al.~\cite{camilo2015bugs} ($\pm$ 0.00 - 0.30: Negligible, $\pm$ 0.30 - 0.50: Low, $\pm$ 0.50 - 0.70: Moderate, $\pm$ 0.70 - 0.90: High, and $\pm$ 0.90 - 1.00: Very high).

\subsubsection{Examining Gender Diversity Changes over Time}
Beyond state of gender diversity based on latest \ghtorrent data, we are also interested in how gender diversity changes over time. 
Considering rapid expansion of \gh in recent years (it has grown from 10 million repositories by end of 2013 to more than 100 million repositories by November 2018~\cite{github2020milestones}), we decide to focus our analyses of change on the period from 2014 onwards.

To create a baseline for comparison, we use the \ghtorrent commit data to identify a set of \gh users who have authored at least one commit to shortlisted projects by 2014. We subsequently apply the same approach used for RQ1 to compute diversity index values for different regions in 2014. We then perform Kruskal-Wallis H test to evaluate the statistical significance of the difference in diversity between 2014 and latest state. Afterwards, we calculate the effect size using Cliff's Delta.

\subsubsection{Examining Gender Diversity of Older versus Newer Accounts}
An additional aspect we are interested in is whether, among commit authors, there is difference in gender balance between older and newer accounts. We investigate this by looking at the account creation years of all commit authors of the shortlisted projects, and compute gender composition for each year between 2014-2018 (the latest year for which \ghtorrent has complete data). 

\subsection{Globally-Distributed Developer Survey}
\subsubsection{Protocol} To understand motivations and challenges faced by developers of different genders in various regions when joining and leaving software projects, we designed and distributed an online survey. The survey  comprised three section of questions. The first section solicits the motivation of developers to contribute, frequency of participation, reasons for selecting a particular project, continue participation, as well as barriers and reasons they have abandoned a software project. We build upon previous surveys on barriers and experiences in online programming communities to develop our survey questions in this section~\cite{lee2017understanding, ford2016paradise, zlotnick2017github}.
To help participants ground their responses, we asked them to answer the above questions for one of the software projects we identified them from.
The second section of our survey included questions about how relevant the gender and region of co-contributors is when selecting a project to contribute to. This section of questions is inspired by how peer parity can encourage participation of people from a shared background or identity~\cite{ford2017someone}.
Relating to region, we ask how challenging it is to contribute with people who speak a different language and the usefulness of translation tools to support that interaction.
Likewise, we asked about the ease of contributing to projects that have contributors with same gender identity and their advice to encourage women participation in GitHub.
Finally, in this section we asked all respondents about what should be done to encourage more women in OSS which is aligned with previous surveys~\cite{canedo2020work, ford2016paradise}. In asking all respondents, we understand better how to approach interventions that not only serve women, but also those of other marginalized identities across geographic regions.
In the third section of our survey, we asked demographic questions about their gender identity and the geographic region they contribute to open source from. All questions were optional and presented as either a Likert scale, multiple-choice, or open response question. The survey was designed to be completed in approximately 7 minutes. 

\subsubsection{Participants} 
We identified survey participants from our \ghtorrent sample.
Our sample comprised all contributors from the selected projects for whom we can infer region, gender, and email address to contact them.
The distribution of contributors was skewed towards some regions (e.g., Northern America was over-represented while Micronesia was underrepresented). 
We observed this skew also in the distribution of men and women across regions.

To gather a representative sample spanning multiple regions, we selected 50 men and 50 women from each region. 
For over-represented groups such as men and Northern America, we randomly identified 50 participants, while for underrepresented groups (with participants less than 50), we selected all contributors.
Overall, we identified 1,562 contributors, of which 1,527 email addresses were valid and did not have an out-of-office reply message. The distribution at region level 2 is shown in Table~\ref{tab:dist_surveyed_authors}, while the total for each region level 1 is shown in Table~\ref{tab:dist_surveyed_authors_level_1}.

\begin{table}[!htb]
\centering
\caption{Distribution of surveyed commit authors at region level 2.}
\label{tab:dist_surveyed_authors}
\begin{tabular}{|l|l|r|r|r|}
  \hline
  \makecell{Region\\Level 1} & \makecell{Region\\Level 2} & M & W & Unknown \\
  \hline
 Africa & Northern Africa & 50 & 3 & 1\\
 Africa & \makecell[tl]{Sub-Saharan Africa} & 50 & 2 & 2 \\
 Americas & \makecell[tl]{Latin America and\\the Caribbean} & 50 & 50 & 17 \\
 Americas & \makecell[tl]{Northern America} & 50 & 50 & 50 \\
 Americas & Others  & 3 & 0 & 0 \\
 Asia & Central Asia  & 22 & 0 & 1\\
 Asia & Eastern Asia  & 50 & 50 & 50\\ 
 Asia & \makecell[tl]{South-eastern Asia} & 50 & 30 & 9\\
 Asia & Southern Asia & 50 & 50 & 15\\
 Asia & Western Asia & 50 & 11 & 5\\ 
 Europe & Eastern Europe & 50 & 49 & 20\\ 
 Europe & Northern Europe & 50 & 50 & 20\\ 
 Europe & Southern Europe & 50 & 39 & 7\\
 Europe & Western Europe & 50 & 50 & 50 \\ 
 Oceania & \makecell[tl]{Australia and\\ New Zealand} & 50 & 39 & 10\\ 
 Oceania & Melanesia & 3 & 0 & 0\\
 Oceania & Polynesia & 4 & 0 & 0\\
 Unknown & Unknown  & 50 & 50 & 50 \\
   \hline
\end{tabular}
\end{table}

\begin{table}[!htb]
\centering
\caption{Distribution of surveyed commit authors at region level 1.}
\label{tab:dist_surveyed_authors_level_1}
\begin{tabular}{|l|r|r|r|}
  \hline
  Region Level 1 & M & W & Unknown \\
  \hline
 Africa & 100 & 5 & 3\\
 Americas & 103 & 100 & 67 \\
 Asia & 222 & 141 & 80 \\
 Europe & 200 & 188 & 97 \\ 
 Oceania & 57 & 39 & 10 \\ 
 Unknown & 50 & 50 & 50 \\
   \hline
\end{tabular}
\end{table}

We received 120 responses (out of 1,527 emails sent; approximately 8\% response rate) in three weeks.
On reviewing the responses, two authors manually analyzed half of the survey responses each for anti-patterns (e.g., all responses are empty or have the same value for all questions). We found two responses with all empty values which we discarded from analysis. We did not observe any other patterns in survey responses. We used 118 responses after discarding the two empty responses.

Our survey garnered approximately one response from a woman (total: 23) for every four responses from men (total: 90). 
Although provided with an option, no participants in our sample identified their gender as non-binary.
Our participants have contributed to open source from around the world, including Europe (46), Asia (29), Americas (21), Africa (12), and Oceania (4), with an overall distribution shown in Table~\ref{tab:dist}.
Some participants preferred not to disclose either gender or geographic region; hence the total count in Table~\ref{tab:dist} is lower than the number of responses received.

\begin{table}[!htb]
\centering
\caption{Distribution of survey responses based on gender and region.}
\label{tab:dist}
\begin{tabular}{|l|r|r|r|}
  \hline
Region	& Men	& Women 	& Total \\ \hline
Europe	& 35 		& 10	 	& 45 \\ 
  Asia 	& 25 		& 4	 	& 29 \\ 
 Americas & 13 		& 7	 	& 20 \\ 
 Africa 	& 11 		& 1	 	& 12 \\ 
 Oceania 	& 3 		& 0	 	& 3 \\ 
 \hline
  Total 	& 87 		& 22	 	& 109 \\ 
   \hline
\end{tabular}
\end{table}

\subsubsection{Analysis}
We had two types of responses: Likert scale and open-ended. To process Likert scale responses, we transformed an ordinal scale into a nominal scale. For example, a 5-point Likert scale of ‘Very important, Important, Neutral, Less important, and Not at all important’ was converted into ‘Important’ (combining ‘Very important’ and ‘Important’ into one), ‘Neutral’, and ‘Not Important’ (combining ‘Less important’ and ‘Not at all important’ into one). This way it is easier to (statistically) distinguish factors deemed important from not important, in addition to the overall distribution. Similarly, other Likert scale questions were processed. 

The transformed nominal scale was fed as input to the Chi-square test to test statistically significant differences in the responses. 
All tests were conducted in R and reported at p$<$0.05. 
For data analysis, we analyze aggregates for which we can draw meaningful inferences. 
Since gendered responses from Oceania are fewer in the count, we remove them from statistical analysis. 

For open response survey questions, the authors conducted a thematic analysis of participant's motivations to contribute, barriers to contribution, and reasons to abandon projects on \gh. 
In the first phase, four authors independently conducted first-cycle descriptive coding~\cite{saldana2009coding} (i.e., summarizing the topic of each response as code) on each open-ended response. In the second phase, one author performed axial coding (i.e., relating the codes to each other) to connect core experiences respondents had in OSS. In the final phase, three authors discussed codes where responses did not converge by negotiation~\cite{campbell2013coding}.

\section{Results} \label{results}
\subsection{RQ1: What are the Gender and Geographic Diversity Characteristics of OSS Projects on GitHub?}
\subsubsection{Regional Variations}
We find that gender diversity of repositories' commit authors are generally low worldwide, as shown in Tables~\ref{tab:gender_diversity_by_region_level_1} and~\ref{tab:gender_diversity_by_region_level_2}. Through Chi-squared test, we found relationship between gender and region (p=6.25e-56 at region level 1 and p=1.30e-78 at region level 2) but negligible association strength (Cramér's V result of 0.07 at region level 1 and 0.08 at region level 2). 

\begin{table}[!htb]
    \caption{Gender diversity (or Blau) index arranged in non-decreasing order by region (level 1). Blau index of 0.5 indicate maximum diversity (50\% men, 50\% women).}
    \label{tab:gender_diversity_by_region_level_1}
    \centering
\begin{tabular}{|l|r|r|r|r|r|}
\hline
Region & Blau index & Commit authors (\% distribution) \\
\hline
Africa & 0.08 & 364 (1\%)\\
Europe & 0.08 & 24350 (34\%) \\
Oceania & 0.10 & 1880 (3\%)\\
Americas & 0.14 & 26607 (38\%)\\
Asia & 0.15 & 5297 (7\%)\\ \hline
Unknown & 0.17 & 12123 (17\%)\\
\hline
\end{tabular}
\end{table}

\setlength\tabcolsep{3pt}
\begin{table}[!htb]
    \caption{Gender diversity index values arranged in non-decreasing order by region (level 2). Blau index of 0.5 indicate maximum diversity (50\% men, 50\% women).}
    \label{tab:gender_diversity_by_region_level_2}
    \centering
\begin{tabular}{|l|l|r|r|r|r|}
\hline
\makecell{Region\\Level 1} & \makecell{Region\\Level 2} & \makecell{Blau\\Index} & \makecell{Commit authors}\\
\hline
 Americas & Others & 0.00 & 5 \\
 Oceania & Melanesia & 0.00 & 5 \\
 Oceania & Polynesia & 0.00 & 5 \\
 Asia & Central Asia & 0.06 & 34 \\
 Europe & Eastern Europe & 0.06 & 3858 \\ 
 Europe & Southern Europe & 0.06 & 2314 \\
 Africa & Sub-Saharan Africa & 0.07 & 273\\ 
 Asia & Western Asia & 0.07 & 529 \\ 
 Europe & Western Europe & 0.08 & 10637 \\ 
 Americas & \makecell[tl]{Latin America and\\the Caribbean} & 0.09 & 2547 \\
 Europe & Northern Europe & 0.10 & 7541 \\
 Oceania & \makecell[tl]{Australia and\\ New Zealand} & 0.10 & 1870 \\ 
 Africa & Northern Africa & 0.11 & 91\\
 Asia & Southern Asia & 0.11 & 1463\\
 Asia & \makecell[tl]{South-eastern Asia} & 0.13 & 686 \\
 Americas & \makecell[tl]{Northern America} & 0.14 & 24055 \\
 Asia & Eastern Asia & 0.20 & 2585 \\ 
 \hline
 Unknown & Unknown & 0.17 & 12123\\ 
 \hline
\end{tabular}
\end{table}
\setlength\tabcolsep{6pt}

The result of our repository-oriented additional analysis at region level 1, shown in Table~\ref{tab:gender_diversity_by_region_level_1_proj}, demonstrates similar ordering from least to most diverse regions. We find that this approach produce overall result that is consistent with result of our previous, region-oriented approach. There is statistically significant difference among regions overall, (p$=$3.89e-115 in Kruskal Wallis H test). We found three pairs with statistically significant difference in Mann-Whitney U test (Americas versus Europe, Asia versus Oceania, and Asia versus Europe, all of which have p$<$0.001). However, we observe negligible effect sizes on Cliff's Delta test ($\delta$ of 0.098 for Americas versus Europe, 0.088 for Asia versus Oceania, and 0.132 for Asia versus Europe).

\begin{table}[!htb]
    \caption{Gender diversity index values by region level 1, computed by associating project with most frequent contributor location.}
    \label{tab:gender_diversity_by_region_level_1_proj}
    \centering
\begin{tabular}{|l|r|r|r|r|r|}
\hline
Region & Mean & Median & Std. dev. & Min & Max \\
\hline
Europe & 0.06 & 0.00 & 0.13 & 0.00 & 0.50\\
Africa & 0.07 & 0.00 & 0.16 & 0.00 & 0.50\\
Oceania & 0.08 & 0.00 & 0.15 & 0.00 & 0.50\\
Americas & 0.09 & 0.00 & 0.16 & 0.00 & 0.50\\
Asia & 0.11 & 0.00 & 0.17 & 0.00 & 0.50\\
 \hline
\end{tabular}
\end{table}

\mybox{
\textbf{Finding}: Gender diversity is low worldwide, and while there is apparent difference in diversity across regions (with Asia and Americas being highest), statistically the difference is not substantial.
}

\subsubsection{Correlation between Geographic and Gender Diversity}
The result of our analysis of correlation between geographic and gender diversity, shown in Table~\ref{tab:gender_and_geographic_diversity_correlation}, shows negligible to small negative correlation between gender diversity and geographic diversity. This suggests that project teams that accept contributors from different regions may still be homogeneous in terms of gender, and vice versa, indicating that different approaches are needed to promote each type of diversity.

\begin{table}[!htb]
    \caption{Spearman's $\rho$ between repositories' gender diversity and geographic diversity. * indicates p-value $<$0.001}%
    \label{tab:gender_and_geographic_diversity_correlation}
    \centering
\begin{tabular}{|l|r|r|}
\hline
Regional Grouping & $\rho$ & p-value \\
\hline
Level 1 (e.g. `Africa') & -0.06 & 0.00* \\
Level 2 (e.g. `Sub-Saharan Africa') & -0.10 & 0.00* \\
Level 3 (e.g. `Eastern Africa') & -0.10 & 0.00* \\
Location (e.g. `Ethiopia') & -0.11 & 0.00* \\
 \hline
\end{tabular}
\end{table}

\mybox{\textbf{Finding}: There is no strong correlation between gender and geographic diversity.
}

\subsubsection{Gender Diversity Changes Over Time}
\begin{figure}[!htb]
\centering
\includegraphics[width=\linewidth]{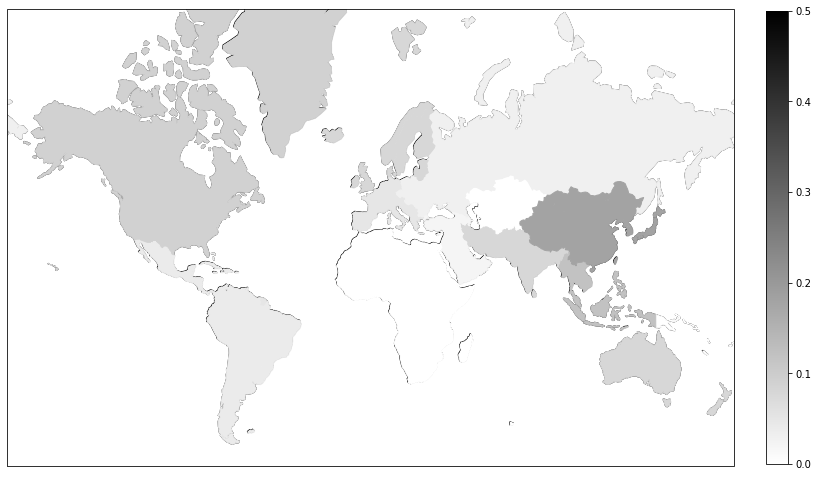}
\caption{Gender diversity at region level 2 as of 2014. Darker shade indicates higher diversity.}
\label{fig:gender_diversity_by_region_map_vis_2014}
\end{figure}

\begin{figure}[!htb]
\centering
\includegraphics[width=\linewidth]{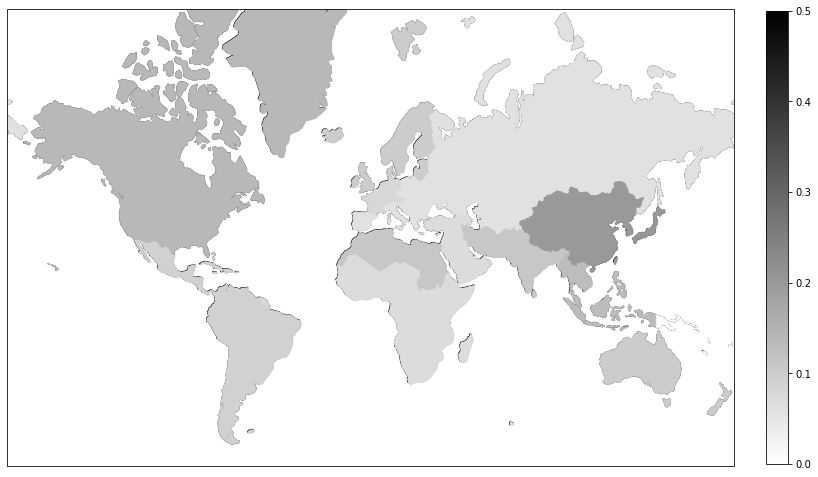}
\caption{Gender diversity at region level 2 as per latest data. Darker shade indicates higher diversity.}
\label{fig:gender_diversity_by_region_map_vis_latest}
\end{figure}

 Table~\ref{tab:rq2_region_level_2_diversity_change} shows the change in Blau index at region level 2, while Figures~\ref{fig:gender_diversity_by_region_map_vis_2014} and~\ref{fig:gender_diversity_by_region_map_vis_latest} show the map visualization. We note that there is general trend of improvement, with most regions showing increase in Blau index value, and none show a decrease. We found that the difference between 2014 Blau index values of the various regions and the latest values is statistically significant (p=0.03), and Cliff's Delta calculation indicate large effect size ($\delta$=0.47). However, as shown in Table~\ref{tab:rq2_region_level_2_diversity_change}, in terms of absolute value, there is still much room for improvement; most regions see an increase in Blau index values of less than 0.10 since 2014, with the exception of Northern Africa, which improved by 0.11.

\setlength\tabcolsep{3pt}
\begin{table}[!htb]
    \caption{Changes in gender diversity of commit authors between 2014 and latest \ghtorrent date - region level 2. N.A. indicates regions for which Blau index cannot be computed since there are no users at the time.}
    \label{tab:rq2_region_level_2_diversity_change}
    \centering
\begin{tabular}{|l|r|r|r|r|r|}
\hline
\multirowcell{2}{Region\\Level 2} & \multicolumn{3}{|c|}{Diversity Index} & \multicolumn{2}{|c|}{Users} \\
\cline{2-6}
& 2014 & Latest & Change & 2014 & Latest\\
\hline
 Northern Africa & 0.00 & 0.11 & 0.11 & 9 & 91 \\
 Sub-Saharan Africa & 0.00 & 0.07 & 0.07 & 55 & 273 \\
 \makecell[tl]{Latin America and\\the Caribbean} & 0.04 & 0.09 & 0.05 & 563 & 2547 \\
 Northern America & 0.09 & 0.14 & 0.05 & 7250 & 24055 \\
 Americas (Others) & N.A. & 0.00 & N.A. & 0 & 5\\
 Central Asia & 0.00 & 0.06 & 0.06 & 6 & 34 \\
 Eastern Asia & 0.18 & 0.20 & 0.02 & 772 & 2585 \\ 
 South-eastern Asia & 0.12 & 0.13 & 0.01 & 159 & 686 \\
 Southern Asia & 0.08 & 0.11 & 0.03 & 207 & 1463 \\
 Western Asia & 0.02 & 0.07 & 0.05 & 113 & 529 \\ 
 Eastern Europe & 0.03 & 0.06 & 0.03 & 962 & 3658 \\ 
 Northern Europe & 0.08 & 0.10 & 0.02 & 2128 & 7541 \\ 
 Southern Europe & 0.05 & 0.06 & 0.01 & 562 & 2314 \\
 Western Europe & 0.05 & 0.08 & 0.03 & 2963 & 10637 \\ 
 \makecell[tl]{Australia and\\New Zealand} & 0.08 & 0.10 & 0.02 & 573 & 1870 \\ 
 Melanesia & 0 & 0.00 & 0.00 & 2 & 5\\
 Polynesia & N.A. & 0.00 & N.A. & 0 & 5\\ 
 Unknown & 0.11 & 0.17 & 0.06 & 2439 & 12123 \\
 \hline
\end{tabular}
\end{table}
\setlength\tabcolsep{6pt}

\mybox{\textbf{Finding}: Globally, the increase in gender diversity in OSS projects is statistically significant with large effect size, however there is still much room for improvement.
}

\subsubsection{Gender Diversity of Older versus Newer Accounts}
\begin{figure}[!htb]
\centering
\includegraphics[width=\columnwidth]{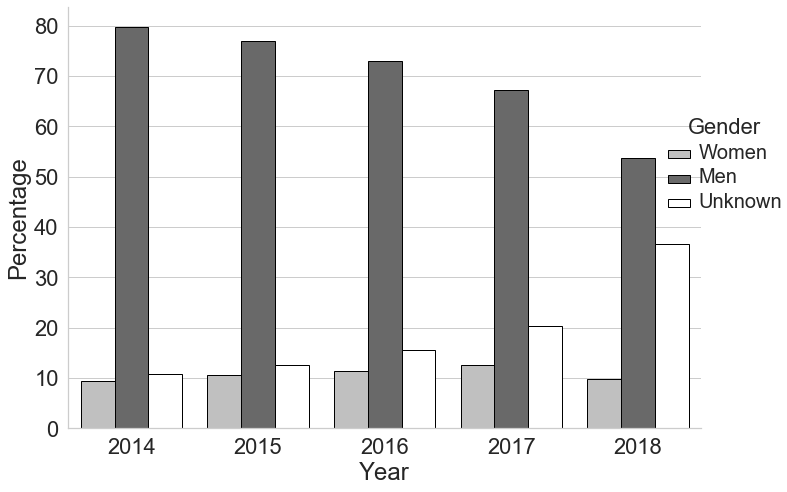}
\caption{Gender percentage of commit authors by account creation year, 2014-2018.}
\label{fig:gender_percentage_by_year_joined_2014_2018}
\end{figure}

Figure~\ref{fig:gender_percentage_by_year_joined_2014_2018} shows the breakdown of commit author accounts by creation year and gender. The percentages indicate that the number of \gh accounts created by women has remained low throughout the period. This suggests a need to encourage participation of women.

\mybox{\textbf{Finding}: Among commit authors with identifiable gender, yearly percentage of account creation by women is around 10\%, suggesting that encouragement of participation is still needed.
}

\subsection{RQ 2: What Factors Potentially Contribute to The Differences in Geographic- and Gender-based Developer Participation?}
We received a range of survey responses from participants that include important factors such as the projects impact, how they are motivated by project alignment, and how they have been inhibited by the community culture. In this section we report the results of our analysis which was done at two levels: globally and regionally. Our objective is to obtain both a global view of factors affecting developer participation, as well as view of any region-specific characteristics that can be utilized to promote participation from particular regions.

\subsubsection{Global Findings}
Overall, we find that the majority of survey respondents contribute to \gh monthly (79), followed by weekly (22), daily (12) and hourly (4) with no differences in contribution pattern across gender and regions.

\noindent\textbf{Project Selection Factors.} 
A majority of developers believe that alignment of project goal to their own is the most important factor for selecting a project. Approximately, 96\% of the respondents consider this factor as important while the remaining 4\% do not consider it important [$\chi^2$ (1 df) = 86.6, p$<$0.001].  
Other factors deemed important are how welcoming the project is (83\% important), how easy it is to join the project (81\%), and the opportunity to be a part of how software is built (79\%). 

Although the majority of participants said they did not select a project because they saw it on social media (94\% not important) or that their friends or colleagues contribute to that project (67\% not important), few acknowledged how other social dynamics did matter. 
For example, some participants mentioned how important it was to them that a project ``supports social equity (P97)'' while providing ``up-to-date code for others learning (P125).''

\mybox{\textbf{Finding}: To encourage participation in a project, goal alignment and creation of welcoming community will be more effective than promotion in social media.}

\noindent\textbf{Motivations To Contribute.} Participants primarily pursued open source software development as their hobby (69 responses), volunteer in the community for free (63), to learn something new (63) or it is their full time job (54) %
Other less prominent reasons are to get a job (22), meet new people (21), as a part of school or university project (8), and to get paid (6). 

From our open responses, participants described their interest in volunteerism as an opportunity to reciprocate what they received from the community in a ``socially relevant (P71)'' way. One participant goes on to say, ``I get so much from the community that I feel where I can I need to give back when I can (P114).''

\noindent\emph{Motivations to Continue Participation.} Once developers have joined a project there are many reasons for developers to continue participation. The factor that is considered most important is interactions with welcoming contributors (91\% of participants consider this important). This is followed by availability of exciting tasks (considered important by 85\%) and the global connections they build worldwide (78\% important).
Low stress level (considered important by 76\%) is another common consideration to continue participation.

\mybox{\textbf{Finding}: While developers may participate in a project for variety of reasons, ensuring continued participation requires project owners to maintain welcoming community, ensuring availability of exciting tasks, and minimizing stress to contributors.
}

\noindent\textbf{Barriers to Contribution.}
From our analysis, we identified 116 barrier statements referring to reasons contributors have decided not participated in some projects or discontinued contributing from others. From these statements we identified 6 themes.

\vspace{.5em}
\noindent\emph{Lack Of Resources.}
Participants acknowledged that they had limited resources at their disposal to make significant contributions to a project. These resources included time allocation, the lack of project funding, and challenges balancing time spent on projects for a full time job with projects a hobbyist. One participant goes on to describe his work-hobby balance: ``I do not do this as a full time job, I just try to commit meaningful changes that helped me in my own projects (P114).’’
Another describes their funding challenges:
``At times I would like to contribute more but it comes down to a lack of funding to put more hours in.  (P112)’’

\vspace{.5em}
\noindent\emph{Goal Alignment Shift.}
As contributors grow in their expertise so do their interests and their professional work. For instance, some participants described how there was a pre-determined end of their ``short-lived project (P26)'', but also that they, ``have abandoned some open source projects because they have been superseded by other projects or because better options for doing the same thing came along (P13).'' Participants did not find useful to stay on a project that was no longer a priority.

\vspace{.5em}
\noindent\emph{Inactivity on Projects.}
Changing project goals often result in projects being abandoned and eventually becoming inactive. 
Participants described the signs of dying project: ``Decrease in the regularity of contributions from project contributors (P70).''
This inactivity on the project went beyond who was contributing. Participants also described significant delay in the code review process from maintainers as a barrier: ``In general, having no frequent experienced contributors would make me stop contributing because reviews from experienced developers is one of my main motives to contribute (P118).'' Contributors are very interested in contributing to projects a as a learning experience, but when the common experience is, ``maintainer just stopped reviewing PRs and abandoned the project (P94),'' contributors lose value in participating.

\vspace{.5em}
\noindent\emph{Poor Engineering Environment.}
Factors related to the engineering environment discouraged contributors. Specifically, participants reported being inhibited by the``complex installation process (P71)'', ``complex code architecture (P70)'', ``lack of documentation (P71)'', and the ``lack of a proper roadmap (P110).'' Without proper documentation and a clear roadmap of what the north star of a project is contributors will be misguided like P79 who had a challenge finding the best opportunities to help: ``On most [projects I'm] not having a clear understanding of what features would be helpful to work on.''

\vspace{.5em}
\noindent\emph{Poor Working Environment.}
Participants disgruntled by their challenges also recalled the toxic work environments some projects can have: ``Sure I have stopped contributing to projects when the maintainers are jerks to me or others.  Other thing that have curtailed or stopped me from working on a project are racism, misogynous behavior or unprofessional conduct by maintainers (P43).'' 
A few participants went on to to discuss their 1:1 encounters with project leadership: ``The big upstream dependency of this project is maintained by a jerk, so I mostly just maintain the project now, rather than actively add new features (P43).''
Although these experiences have been described in low frequency, it is important to note that these experiences can influence how developers decide to contribute like in P43's case.

\vspace{.5em}
\noindent\emph{Unclear Onboarding.}
The lack of official onboarding documentation processes from maintainers was also discouraging to our participants: ``My contribution there was very small, as we did not use it a lot.  But I guess this is a good example of the not very well documented project. this is the main obstacle for me when I would like to get involved in some project - not very clear README, missing documentation regarding code discipline for a particular project, not clear rules on how to get involved. That would be for me the main blocker (P98).'' 
When participants reflected on their past experiences with their first project they recalled how challenging it was to join some projects: ``The first contact is always the hardest, I mean the totally new newbies always find it intimidating to find and join their first project. (P95)''
In short, new contributors to a project have a hard time finding how to get involved.

\subsubsection{Gender and Regional Related Motivations and Challenges}
We found that women developers place high importance on social aspects related to OSS projects as an aspect to consider before participating. Women value selecting a project with friends and colleagues more than men (64\% of women participants consider this important, compared to only 25\% of men). Beyond this, 37\% of women developers believe that shared gender identity with fellow contributors as important, while only 1\% of men consider it important. Analysis across regions showed that same gender identity is not at all important for developers from Africa (0\%) while it does hold some relevance for other regions: Americas (17\%), Europe (11\%), and Asia (4\%). Beyond the social aspect, we also found that being paid is a greater incentive for women (64\% find it important) compared to men (35\% find it important).

We also asked participants about what they think can encourage participation among women on \gh. We found that some men across regions were very dismissive to this question saying, ``Ask the women. I'm not stopping them (P9).'' On the opposition, we also did find some men suggesting how explicit visibility can inspire others, ``There were several women highly qualified for any type of project. But if you need any encouragement, perhaps more women will take the initiative to start new open source projects. Maybe it's contagious (P26).'' Likewise, we find that most women were interested in women encouraging other women, but through leadership: ``More women reviewers. More women acting directly on the governance of large open source projects (P52).''
Additional details about this finding can be found in Appendix B%

\mybox{
\textbf{Finding}: 
Shared gender identity, working with friends and colleagues, and being paid is more important for women than men. 
}

\subsubsection{Regional Variation in Motivations and Challenges}
\noindent\textbf{Motivation to Participate in OSS Projects.}
Table~\ref{tab:motivationRegion} shows the developers' motivation to participate in OSS projects, broken down by region. %
We find that the  motivation to contribute to OSS as a full-time job is less common outside of Europe and the Americas. In addition, developers from Africa place a relatively higher importance on networking (i.e., ``meeting new people'') compared to developers from other regions.

\begin{table}[!htb]
\centering
\setlength\tabcolsep{3pt}
\caption{Motivation of developers to participate in open source software projects across regions. Each cell 
reports the percentage of developers motivated by the following factors.}
\label{tab:motivationRegion}
\begin{tabular}{|l|r|r|r|r|}
  \hline
 & Europe & Asia & Americas & Africa \\ 
  \hline
my full-time job & 26.00 & 11.00 & 21.00 & 8.00 \\ 
  my hobby & 21.00 & 28.00 & 15.00 & 19.00 \\ 
  volunteer for free & 26.00 & 20.00 & 17.00 & 22.00 \\ 
  learn something new & 15.00 & 24.00 & 25.00 & 22.00 \\ 
  \makecell[l]{school/university project} & 2.00 & 1.00 & 8.00 & 0.00 \\ 
  help get a job & 3.00 & 8.00 & 8.00 & 11.00 \\ 
  meet new people & 5.00 & 6.00 & 6.00 & 14.00 \\ 
  get paid & 2.00 & 1.00 & 0.00 & 3.00 \\ 
   \hline
\end{tabular}
\end{table}
\setlength\tabcolsep{6pt}

\noindent\textbf{Motivation to Continue Participation in OSS Projects.} Table~\ref{tab:motivationRegion2} shows the developers' motivation to continue their participation in OSS projects, broken down by region. We note that there are regional variations regarding importance of various factors. For example, while exciting and challenging tasks are important for all regions, they are more important for developers from Asia and Africa. On the other hand, connecting with people worldwide is not a big motivation for developers from Europe and Americas to continue participation.

We also found regional differences between what motivates developers to \textbf{participate} and what motivates developers to \textbf{continue} participation. This difference is in line with Gerosa et al.'s finding~\cite{gerosa2021shifting} regarding shift in motivation of OSS contributors as these contributors gain tenure. For instance, as shown in Table~\ref{tab:motivationRegion}, the percentage of African developers who participate in OSS as full-time job, to help get a job, or to get paid is relatively small. However, Table~\ref{tab:motivationRegion2} shows that being paid is an important consideration for African developers to continue participation, much more so than it is for developers from Europe, Asia, and America. This suggests that while African developers may start participating in OSS projects as a hobby, to volunteer, or to learn something new, monetary rewards are important to maintain long-term participation. As another example, while a small percentage of Asian developers stated ``meeting new people'' as reason to participate in OSS projects, 89\% reported connecting with people worldwide as a reason to continue participation---a percentage similar to developers in Africa (86\%).  %

\begin{table}[!htb]
\centering
\caption{Reasons to continue participation in open source software projects across regions. Each cell reports the percentage of developers that find the following factors important or not important.}
\label{tab:motivationRegion2}
\begin{tabular}{|l|r|r|r|r|}
  \hline
 & Europe & Asia & Americas & Africa \\ 
  \hline
\multicolumn{5}{|c|}{Interactions with welcoming contributors} \\
\hline
Important & 86 & 96 & 94& 100 \\
Not important & 14 & 4 & 6 & 0 \\ \hline
\multicolumn{5}{|c|}{Connects with people worldwide} \\
\hline
Important & 67 & 89 & 77 & 86\\
Not important & 32 & 11 & 23 & 14 \\ \hline
\multicolumn{5}{|c|}{Exciting tasks} \\
\hline
Important & 75 & 100 & 77 & 92\\
Not important & 25 & 0 & 23 & 8\\ \hline
\multicolumn{5}{|c|}{Challenging tasks} \\
\hline
Important & 84 & 100 & 82 & 100\\
Not important & 16 & 0 & 18 & 0 \\ \hline
\multicolumn{5}{|c|}{Being paid} \\
\hline
Important & 34 & 38 & 21 & 71 \\
Not important & 66 & 62 & 79 & 29\\ 
   \hline
\end{tabular}
\end{table}

\mybox{
\textbf{Finding}: Some form of funding for participation in OSS projects can be particularly effective to promote continued participation of developers from Africa.
}

\noindent\textbf{Relevance of Shared Regional and Linguistic Identity.} 
Overall, having contributors from same geographic region in the project is not important for contribution, albeit subtle differences exist across regions. Having contributors from the same geographic region is least important for Europe, followed by Americas, Asia and somewhat important for the developers from Africa (see Table~\ref{tab:sameRegion} for details).

We also solicited challenges in working with people who speak a different language, and noticed that while overall differences are not discernible, at regional level, the responses are quite divided. 
Developers from Europe who happen to see no value in having contributors from same region also do not find it challenging working with developers who speak a different language. 
Developers from Africa, on the other hand, not only find it relatively more important to have fellow developers from the same region in the project, but also have difficulty in interacting with contributors who speak a language different from theirs. Meanwhile, developers in Asia and America are evenly split in their responses (see Table~\ref{tab:sameRegion} for details). We also found that developers overall hold mixed opinion on the usefulness of translation tools, with no differences across regions. However, there is a difference across genders. We found that 76\% of women developers find translation tools helpful, but only 55\% of men developers do so. 

\begin{table}[ht]
\centering
\caption{Relevance of shared regional identity and language across geographic regions.}
\label{tab:sameRegion}
\begin{tabular}{|l|r|r|r|r|}
\hline
& Europe & Asia & Americas & Africa \\ \hline
\multicolumn{5}{|c|}{Contributors from same geographic region} \\
\hline
Important 		& 9 & 19 & 15 & 40 \\
Not important 	& 91 & 81 & 85 & 60 \\ \hline
\multicolumn{5}{|c|}{Working with people who speak a different language} \\
\hline
Challenging & 26 & 50 & 50 & 80 \\ 
Not challenging & 74 & 50 & 50 & 20 \\ 
   \hline
\end{tabular}
\end{table}

\mybox{
\textbf{Finding}: Provision of better translation tools will be particularly helpful to encourage participation of women developers worldwide, as well as participation of developers from Africa.
}

\section{Discussion} \label{design_implications}
\subsection{Summary of Findings}
Our result for RQ1 did not show substantial difference across different geographic regions. We note that the set of commit authors with unresolved location has higher apparent Blau index compared to sets from known regions. A factor that contributes to this is the high percentage of users in the set whose gender is also unresolved (31.82\%, as shown in Table~\ref{tab:overview_developers}). Since the Blau index calculation ignores ``Unknown'' gender, and majority of commit authors are probably men (based on proportions of commit authors whose gender and location can be resolved), we believe the high percentage of unknowns increases apparent women-to-men ratio in favor of women. This subsequently increases the Blau index of the group with unknown location.

As for the observed diversity improvement during the period analyzed in this work, we believe it is influenced by a combination of factors. Firstly, in recent years there has been increasing interest in promotion of diversity in computing. This includes efforts by non-profit organizations (such as Girls Who Code\footnote{https://girlswhocode.com/}, Women Who Code~\footnote{https://www.womenwhocode.com/}, NCWIT\footnote{https://www.ncwit.org/}, and ACM-Women~\footnote{https://women.acm.org/}), programs targeted at school students~\cite{ibe2018reflections, vachovsky2016toward}, initiatives by universities to improve diversity in their own programs~\cite{borsotti2018sigsoft,  kulkarni2018promoting, rheingans2018model}, as well as efforts by various organizations worldwide to hire more diverse staff. This occurs along the growth of the software industry including in previously underrepresented regions such as Africa~\cite{kelly2016tech}, with GitHub itself seeing a drastic increase in popularity outside the United States\footnote{https://github.blog/2018-11-08-100m-repos/}. These factors help attract more diverse talents into computing, including women from underrepresented regions. Nevertheless, as the data shows, there is still much room for improvement.

Related to RQ2, survey responses from our participants encourage us to consider what mechanisms can support contributors from specific regions. In summary, our findings highlight three approaches that should be utilized to better support inclusion across gender and geographic regions. They are:
\begin{enumerate}
    \item Development of friendlier communities, especially towards newcomers.
    \item Highlighting of role models from marginalized communities.
    \item Augmentation of existing automated software engineering techniques to incorporate social factors.
\end{enumerate}

\subsection{Opportunities Ahead}
\subsubsection{Development of Friendlier Communities}
There are several ways to encourage development of friendlier, more welcoming communities. Creation and enforcement of codes of conduct are an example of a way to promote a safe environment that can support inclusion~\cite{singh2019women,ehmke2014covenant, ford2019remote}. 
Having a code of conduct can support a two-pronged approach of: 1) allowing lurkers interested in contributing (e.g., including women and other marginalized developers) to feel more comfortable in contributing since they know there are guidelines that can protect them from toxic interactions and 2) signal to developers who are already in the community (e.g., including those that may have been inciting toxic interactions) that there will be repercussions for their actions. 
Unfortunately, less than 10\% of the top OSS projects actually have one~\cite{singh2019open}. Participants in our survey also acknowledged that one thing that would encourage inclusion is ``Promoting use of and enforcement of code of conduct (P94).'' Even fewer projects are transparent about how they enforce these guidelines, if at all.

One approach to enforcing code of conduct usage is rewarding projects that have one. For example, \gh can offer donation through sponsors program as a reward for projects that have code of conduct. This will provide maintainers with more resources to devote to their role, encourage them to make sure their project is inclusive, and signal to new contributors that a project is safe. Comparatively, this presents a missed opportunity by the projects that have not provided an enforceable code of conduct and thus incentivize those projects to adhere to a new norm. A risk of this approach is the possibility of project maintainers creating token codes of conduct just to satisfy conditions to receive rewards. This approach should therefore be coupled with evaluation of the code of conduct to ensure that it is both meaningful and actually enforced.

Beyond code of conduct, other potential ways to promote development of friendlier communities are usage of social metrics for community self-evaluation and improvement. A example may be drawn from sites that show employer reviews such as GlassDoor\footnote{https://www.glassdoor.com/} and various job search portals. In OSS context, ability to provide and show contributor reviews as well as other metrics such as distribution of contributor tenure can help developers evaluate potential projects to join, and also provide an OSS project community a means to evaluate what they have or have not done well and how to improve their community.

\mybox{
\noindent\textbf{Challenge:} Many communities currently do not have or enforce code of conduct, and aspiring contributors also can't easily evaluate community quality of a given OSS project.\\
\noindent\textbf{Opportunity:} Improvements can be done by promoting creation and usage of codes of conduct across communities, and to provide set of social metrics to help aspiring contributors evaluate quality of community they consider joining.
}

\subsubsection{Mentorship and Highlighting of Role Models}

\noindent\textbf{Highlighting of Regional / Women Developers as Role Models}. 
From the responses, contributors from underrepresented OSS regions are not necessarily resentful. Rather, they would like to empower people from their region to take part in the opportunity to be a builder of software that people around the world use~\cite{brodsky2018chinese,abati2017nigeria}. One participant from Sub-Saharan Africa went as far as to state 
``Open-source software is a solution for Africa to progress as a continent as quickly as possible while spending less money (P23)''.

To support and further activate opportunities such as these, we propose a proximity-based mentorship where mentors and mentees are relatively close in region or even close in cultural dimension (e.g., survival vs. self expression~\cite{oliveira2016participation}). This experience can take advantage of being in the same shared region by conducting guidance through offline interventions~\cite{debian2019mentoring}. The duality of fostering both the same community online based on a personal offline experience can further support inclusion.

Another approach that can be used is to highlight role models from underrepresented demographics. For example, our survey results indicate that women developers are interested in mechanisms that highlight the contribution of women. Such mechanisms can be implemented both online and offline. Online mechanisms can be in the form of updates to pages such as \gh Explore~\cite{github-explore} to add sections that highlight rising or top developers from underrepresented communities. For offline implementation of this mechanism, developer communities can for example organize and encourage technical presentations and talks by experienced developers from underrepresented demographics.

\mybox{
\noindent\textbf{Challenge:} There is lack of mechanism to highlight contribution of developers from underrepresented demographics.\\
\noindent\textbf{Opportunity:} Mechanisms that highlight developers that are popular globally can be augmented to also highlight top or popular developers from more specific demographics.
}

\subsubsection{Diversity Promotion via Automated Software Engineering Tools}
Some barriers appear to present opportunities for applying automated software engineering approaches to attract diverse contributors to OSS projects. Existing works~\cite{casalnuovo2015developer, hahn2008emergence} highlight the importance of prior social links with existing contributors in developers' decision to join an OSS project, and this can be exploited to promote diversity by augmenting existing approaches with social considerations. We discuss some specific categories of tools in the following paragraphs.

\noindent\textbf{Automated project recommenders} can be augmented to take into account social considerations. %
A small number of recent project recommenders~\cite{liu2018recommending, matek2016github} factor in developer's social ties, and \gh itself takes into account which developers a user ``follows'' when recommending projects in its \gh Explore~\cite{github-explore} page. However, to promote diversity or participation from particular gender/region, these can be further augmented with additional metrics based on recommendations in the survey responses, for example:
\begin{itemize}
    \item Metrics related to quality of community. For example, typical tenure of contributors (as a proxy of how much contributors enjoy being in the community), reputation of current contributors, and range of current contributors' experience levels (as a proxy of how welcoming the project is to beginners).
    \item Number of current contributors known to be from similar region as the developer considering to join the project.
    \item Diversity of current set of active contributors with known gender and/or location.
\end{itemize}

\noindent\textbf{Automated documentation improvement} can be employed more widely to reduce barriers to contribution. This can include application and enhancement of automated document localization  techniques  to  overcome  language  barriers  and support  local  languages  from  regions  with  large  numbers of  potential  contributors. This may be coupled with application of automated techniques to improve readability, completeness and/or quality of artifacts such as README files~\cite{prana2019categorizing} and release notes~\cite{moreno2014automatic}. Usage of automated document generation of source code summary~\cite{mcburney2014automatic} and tracking of outdated API names~\cite{lee2019automatic} can further reduce time required from potential contributors. This will be valuable especially in regions where OSS projects are more commonly treated as hobby or volunteer work, since reduced time barrier will enable more people to contribute even without monetary rewards.

\noindent\textbf{Automated developer assignment} mechanisms can be updated to distribute exciting / challenging tasks more widely to motivate continued participation. This may be in form of modification to existing automated bug assignment techniques such as~\cite{jonsson2016automated} and~\cite{ zhang2017bug}, that currently are usually used to speed up resolution process~\cite{zou2018practitioners} instead of to spreading interesting tasks to team members.

\mybox{
\noindent\textbf{Challenge:} Current automated software engineering tools tend to focus on technical aspects and similarity between developers (homophily) when making recommendations.

\noindent\textbf{Opportunity:} There's opportunity to augment existing tools to enable selection of target social objectives, such as maintenance of contributor interest (by making more even distribution of challenging tasks) or encouraging participation from certain underrepresented communities.
}

\section{Threats to Validity} \label{threats_to_validity}
\noindent\textbf{Construct Validity.}
Our study has two parts: a large scale data analysis and a survey.
During the study design, we made choices that can potentially influence the outcome. Regarding repository selection, the filtering criteria we use still leaves some possibility of including repositories of academic projects that run beyond 6 months, however, we believe that those are also likely to be a more serious endeavor instead of simple programming assignments.
Another factor is the accuracy of gender and location resolution.
While many factors can cause incorrect gender and location resolution (e.g., incorrect information on GitHub profile, decision to make accounts private), we tried mitigating this threat in two ways. First, we choose a tool that has reportedly reasonable accuracy for multiple regions such as Asia and Eastern Europe~\cite{santamaria2018comparison,karimi2016inferring} and has been used in various studies related to gender representation~\cite{holman2018gender, thomas2019gender, russell2018large}. Prior to full-scale analysis, we also performed validation by manually checking a subset of the data to increase our confidence in the gender prediction. We also limited our analysis to commit authors, who are more likely to be a code-contributing part of the project team (compared to, for example, issue reporters) and are also more likely to provide information which can be used to resolve their gender and location. Finally, we eliminated projects for whom we could not infer gender and location of at least 75\% of commit authors. While it is also possible to perform additional validation after the survey by comparing  self-reported gender and geography in the response to the information inferred from data analysis, we did not do so as we did not ask prior permission from survey participants for such data usage. This is in compliance with the GDPR and broader research ethical considerations.

We also note that the tool we use (\textit{genderize.io}) is not reflective of a broad gender spectrum. While analysis of non-binary identities is a research challenge that has received increasing research attention~\cite{hamidi2018gender,keyes2018misgendering}, we are currently unaware of methods to reliably assess this in software systems at a large scale. Future research should investigate this deeper. As none of our survey respondents identified themselves as non-binary, we believe this limitation of genderize.io does not propose a significant threat to the validity of our subsequent analyses.

With respect to our survey, the underrepresentation of women and a broader set of commit authors poses a threat to validity.
We attempted to mitigate this by using stratified survey sampling based on gender and location, instead of performing a random sampling of the entire population. For focused survey responses, we asked each participant questions relating to a specific project which we hope provide more concrete response based on the participant's own experience, although there is still some validity risk if the participant has not worked on the project recently. 

\noindent\textbf{Internal Validity.} 
Our analysis indicates regional and gender-based differences for open source participants on GitHub.
To improve the internal validity of our data analysis, we calculated diversity at different times using two metrics.
Our results point in the same direction. Likewise, our survey borrows elements from literature (corroborating with its findings) and builds on it. Using strategic sampling techniques we tried to gather a representative sample to offer a worldwide view.

\vspace{.5em}
\noindent\textbf{External Validity.} 
The representatives of our findings is defined by the range of software projects studied.
We selected a wide variety of software projects, nevertheless, we might have systematically missed projects which did not meet our prerequisites (e.g., infer gender and location). 

Likewise, due to our methodology and scope of respondents at the intersection of both marginalized genders and underrepresented countries in OSS, we miss the opportunity to provide broad insight into the challenges of having an intersectional identity~\cite{rankin2019straighten}. Further intersectional methodologies and frameworks should be adopted to explore and amplify the voices of developers in the margins.

\vspace{.5em}

\section{Conclusion and Future Work} \label{conclusion}
In this paper, we report findings from our large scale empirical study leveraging quantitative data from \gh and qualitative data for a targeted survey to developers to report on the gender differences across geographies. Our study finds that there is low diversity across regions worldwide, and although there is some variation among regional diversity, the difference is not substantial. Since 2014, there has been small and statistically significant improvement of gender diversity amongst software contributors in North America and South-Eastern Asia but negligible change elsewhere. We observe that among commit authors with identifiable gender, yearly percentage of account creation by women remains low. A qualitative analysis shows that many of the barriers and motivations for contributing converge across different geographic regions ranging from lack of resources, goal alignment shift to poor working environments and unclear on boarding.

There are two underlying themes we hope this study will achieve. The first is quantifying and setting baseline of current state of \gh regarding intersection of gender and geography. This will help other researchers build on it and quantify changes in coming years. The second is to create awareness of this problem and hopefully encourage further research by the community towards reducing the gender gap and make software contributions possible by everyone, everywhere. Towards this goal, we are working with people in \gh and Stack Overflow to help drive some of the concrete observations from our study to alleviate diversity-related issues in the coming years.

Finally, we also believe it will be helpful if researchers from the different parts of the world perform more in-depth study of gender differences in their own regions. We believe that with better understanding of and connections with local developer communities (including developers who are not active on \gh), local researchers will likely be able to collect more responses. Further, they will also be able to customize their survey to better focus on any region-specific issues they are aware of.

\section*{Acknowledgements}
We would like to thank our survey respondents for providing their perspectives for this study. This research is supported by and the Research Lab for Intelligent Software Engineering (RISE) Operational Fund from the School of Computing and Information Systems (SCIS) at Singapore Management University (SMU).

\section*{Dataset Availability} \label{dataset_availability}
In the interest of encouraging others to replicate and build upon our work, we are sharing our data. The data for this study can be found at: 
\href{https://doi.org/10.5281/zenodo.4637095}{\includegraphics[width=3cm,trim=0 1mm 0 0]{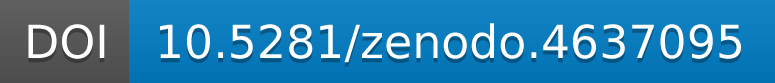}}

\appendices
\begin{landscape}
\section{Literature Table}~\label{sec:lit-table}

\begin{table}[!htbp]
\centering
\caption{Comparing the research goals, research methods, and key findings from closely related literature on gender and geographic diversity.}
\label{tab:relatedwork}
\renewcommand{\arraystretch}{1.5} 
\begin{tabularx}{0.86\linewidth}{
|>{\raggedright}p{0.080\linewidth}
|>{\raggedright}p{0.135\linewidth}
|>{\raggedright}p{0.135\linewidth}
|>{\raggedright}p{0.135\linewidth}
|>{\raggedright}p{0.135\linewidth}
|>{\raggedright\arraybackslash}p{0.135\linewidth}|}
\hline
& 
\textbf{Geographic, Gender\newline Inclusion (This Paper)} & 
\textbf{The Shifting Sands\newline of Motivation\newline 
(Gerosa et al., 2021)~\cite{gerosa2021shifting}}& 
\textbf{Women Core Developers\newline (Canedo et al., 2020)~\cite{canedo2020work}}& 
\textbf{Diversity, Where\newline Do We Stand\newline  (Bosu et al., 2019)~\cite{bosu2019diversity}}& 
\textbf{Diversity Teams\newline Study\newline (Ortu et al., 2017)~\cite{ortu2017diverse}}\\ 
\hline
{\textbf{Research questions/ goals}} & 
{Identify how gender diversity has changed over time and across regions. Identify gender-based and geographic-based developer participation}& 
{Identify what motivates OSS contributors and how contributors' motivations change as OSS matured (e.g., contributors themselves gaining tenure)} & 
{Identify the gender diversity and work practices of core developers in OSS communities} & 
{Determine the level of gender diversity among popular OSS projects}& 
{Understand the impact of gender and nationality diversity on team productivity and collaboration quality}\\ 
\hline
{\textbf{Research methods}} & 
{Mixed-methods study of GitHub Projects over time and conducting a purposefully sampled survey with developers across geographic regions and identifiable genders (men, women and unidentifiable)} & 
{Survey of OSS contributors recruited from social media sites (e.g., Twitter, Facebook, Reddit, LinkedIn, and Hackernews}, through groups related to OSS development, and personal contacts & 
{Mixed-methods study of mining software repositories, identifying gender of contributors and interviewing women core developers} & 
{Mined code review repositories of the top 10 popular OSS project on GitHub} & 
{Built regression models comparing collaboration in issues and dialogue of politeness}\\ 
\hline
{\textbf{Population/ initial sample}} & 
{125,485,095 projects from GH Torrent} & 
{Open to all self-reported OSS contributors. Filtered on reported experience and response validity} & 
{Top 100 most popular projects written in the top 15 most popular programming languages} & 
{Top 10 OSS Projects using Gerrit and had at least 15,000 code reviews} & 
{2014 GHTorrent dataset scoped to closed issues with 2 comments}\\ 
\hline
{\textbf{Participants/ data}}& 
{21,456 repositories, 70,621 commit authors, 122 survey respondents across 5 large geographic regions and across genders} & 
{242 responses from 5 different continents. Includes 82\% men, 81\% coders, 26\% report being paid for contributions} &
{711 projects, 35 women core developers} &
{683,865 pull requests, 4543 non-casual contributors} &
{33,673 issues with 71,423 comments posted by 13,872 developers}\\ 
\hline
{\textbf{Findings}}& 
{- Noticeable differences across gender diversity in regions specifically, Asia and Americas being the highest\newline
- Barriers and motivations to contributing converge across geographic regions\newline 
- No strong correlation between gender and geographic diversity}&
{- Main motivations differ between novice contributors and experienced ones\newline
- Motivations such as learning and knowledge sharing remained important overtime. While altruism increased in importance and self-serving usage decreased in importance.}&
{- OSS has horizontal and vertical segregation\newline
- No significant differences between work practices of men v. women}&
{- Women make up less than 10\% of core contributors\newline
- No significant difference among men vs. women on selected projects}&
{- Higher gender diversity$\rightarrow$ lower team average issue fixing time\newline
-  Nationality diversity $\rightarrow$ lower team politeness}\\ 
\hline
\end{tabularx}
\end{table}

\end{landscape}

\section{Survey Results: Encouraging Women}
~\label{sec:encourage-women}

We asked all survey respondents (regardless of identified gender) about what they think can encourage more women in \gh and received 73 responses to this questions. We qualitatively analyzed responses to this question via open coding and axial coding process which included an iterative review of our themes.
We grouped response according to regions, but despite using our strategic sampling approach, we did not have a critical mass of responses across regions and genders to make meaningful conclusions. We provide summaries of responses by the following themes below supplemented with quotes.

\subsection{Encouragement through awareness} %
We received several broader responses about how to encourage women through providing awareness via several activities. 
\blockquote{Awareness will go a long way in encouraging women to participate. A lot of people would love to participate but wasn't sure where to start.}{14}

One activity mentioned was making education more accessible:
\blockquote{Accessible education. I think many women and girls don't realise they have the skills needed to contribute to programming \& software projects. Teaching young women that they have the potential to do this is really important.}{13}

We also identified several responses about fostering a more welcoming community that included fewer misogynistic developers and more overall encouragement:
\blockquote{Fewer misogynistic developers.}{25}
\blockquote{De-stigmatize programming as a male dominated profession.}{86}

Although strategies in this category were rather general, they indicated that respondents are familiar with challenges some women face in OSS communities.

\subsection{Creating opportunities}
Respondents also indicated concrete recommendations on what may support encouraging women on \gh. Some of these recommendations include adding events that specifically support women, such as having women-centered events or even amplifying the presence of women that are active in the community:
\blockquote{A women support circle is nice, I've seen effort and took part in some but I find women are more comfortable and more encouraged within the same gender group. }{17}
\blockquote{Show what women who are working with this are doing and how is their experience, do projects/workshops.}{35}

\blockquote{More women reviewers. More women acting directly on the governance of large open source projects.}{52}

Likewise other participants mentioned that the community should emphasize the use of existing mechanisms, such as project codes of conduct:
\blockquote{Enforce codes of conduct.}{79}
\blockquote{more welcoming in projects, a well-defined code of conduct to make them feel more comfortable.}{101}
\blockquote{Giving more visibility and fighting against bad behaviors by other men.}{87}

Participants also cite that encouraging women to be apart of 'open developer sprints' can provide more clarity on multi-phase contribution processes:
\blockquote{Encourage more open devsprints and workshops to help women get started easily. More hands-on sessions on upstream contributions.}{37}

\blockquote{Creating awareness among women contributors and building confidence, by conducting interactive sessions on open source and contributions.}{47}

Other broader recommendations participants made were to fund developers making contributions,\inlinequote{Getting paid}{19}, and being more transparent in the code review process, \inlinequote{being more articulate about feelings and motivation behind some critique that could come up for example in code review}{51}.

\subsection{Outside of GitHub}
Many respondents reported that the solution to getting more women to engage in OSS is an issue that goes broader than \gh. Some participants reported that there should be a focus beyond computer science and on STEM fields in general:
\blockquote{I guess you need help more women go to colleges and learn STEM, and make sure they will not be rejected from some professional jobs in the STEM field after graduation.}{96}

\blockquote{We just need more women in programming overall, and I think school outreach programs are the best thing.}{64}

Another set of respondents indicated that changing the bro-culture of technology in the software world is the approach to take: 
\blockquote{Don't see why Github has anything to do with women's participation. In general, low participation to oss from women's may be related to they being minority in the whole bro cultured software world.}{88}
\blockquote{Men should know how to interact with women without an air of authority and welcoming...}{31}

Some respondents described that there are broader global issues that persist outside of work:
\blockquote{Global women's rights, not in IT only.}{34}
\blockquote{Treating women as equals.}{33}

These findings indicate that respondents were aware of challenges in tech but also it made sense to address issues at a wider scale. 

\subsection{The `I Don't Care' Responses}
Finally, we did receive several responses that a) either dismissed questions that focused on the experiences of women in OSS, b) were unclear on the challenges that women face, or c) actively responded with a negative tone to this question (as opposed to simply leaving this optional question blank). As we have not seen previous literature acknowledge this negative sentiment towards empowering a marginalized group, we found it imperative to share the responses we received here:

\blockquote{It is naturally that women are less interesting in technologies than men. I don't see any barrier to prevent women to participating in Github.}{109}

\blockquote{I don't know why this is even a thing.}{41}

\blockquote{Ask the women. I'm not stopping them.}{9}

We also had several toxic recommendations suggesting women look into \inlinequote{Sex reassignment surgery}{53} and that \inlinequote{Good engineers should help themselves}{24}. We highlight these responses not to amplify these biased perspectives, but to show that there are OSS contributors in the community who \emph{do not} understand that there is an issue with the gender diversity.
It is not the job of the marginalized contributors to `fix' the community---it is up to \emph{everyone} to create an inclusive environment. Future work should explore interventions that create a broader awareness of why it is important for everyone to be inclusive along gender and regional diversity.

\smallskip
We hope that these responses encourage researchers to study a variety of gender experiences (including non-binary genders) to capture rich-region specific diversity issues. 
Having more region-specific studies will allow us to provide bespoke solutions that take into the cultural nuance of each region.

\bibliographystyle{IEEEtran}
\bibliography{main}

\begin{IEEEbiography}[{\includegraphics[width=1in,height=2.5in,clip,keepaspectratio]{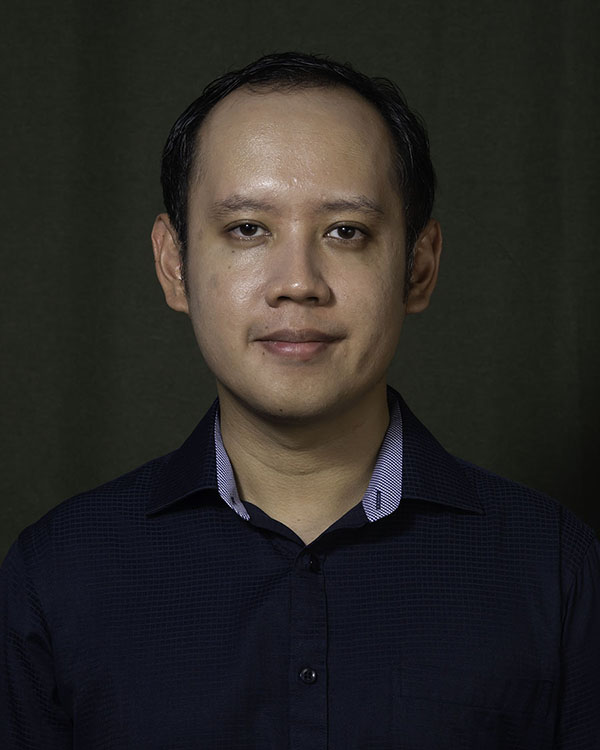}}]{Gede Artha Azriadi Prana} is a PhD student in the School of Computing and Information
Systems, Singapore Management University, under supervision
of Professor David Lo. His research focuses on software engineering analytics. Prior to his current study, he worked in the field of software engineering for about a decade.
\end{IEEEbiography}
\vskip -1.5 \baselineskip plus -1fil
\begin{IEEEbiography}[{\includegraphics[width=1in,height=2.5in,clip,keepaspectratio]{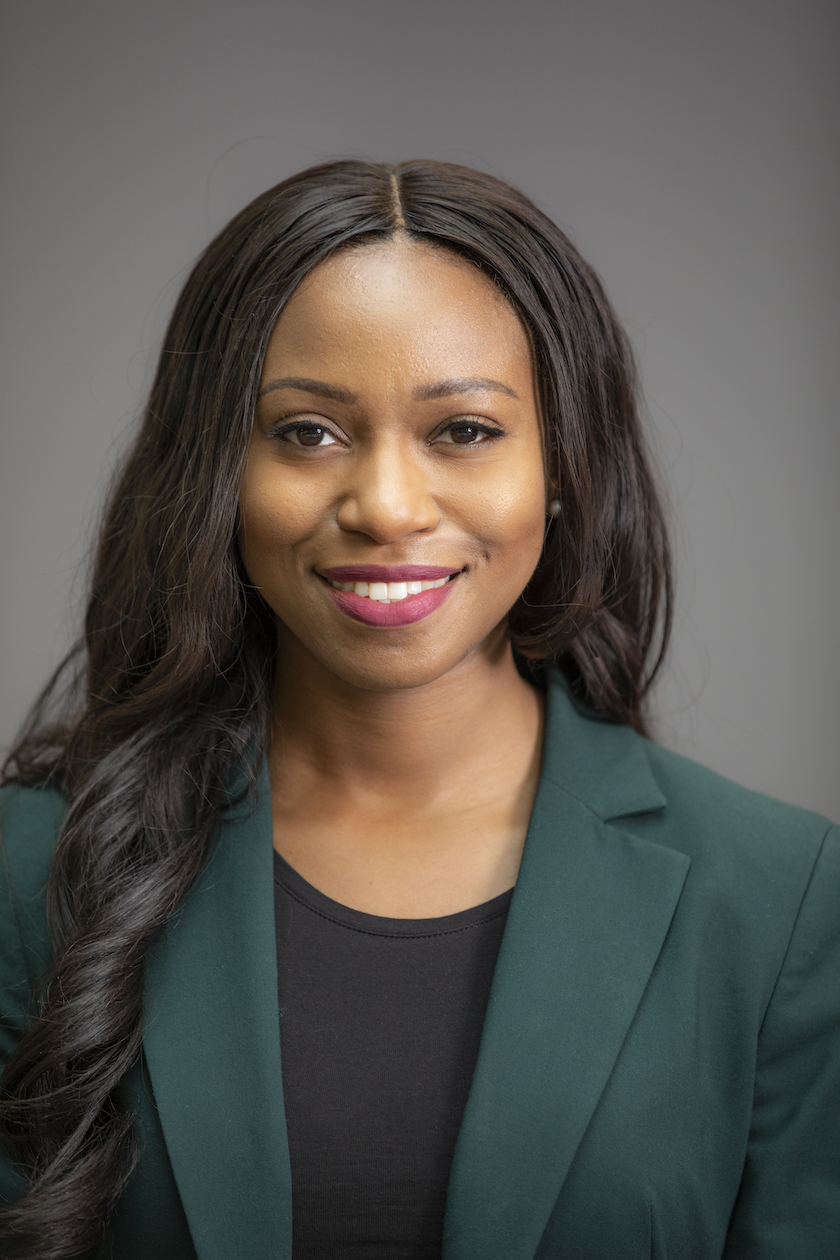}}]{Denae Ford} is a Senior Researcher at Microsoft Research where she works in the Software Analysis and Intelligence (SAINT) team and an Affiliate Assistant Professor at the University of Washington in the Human-Centered Design and Engineering Department. She received her Ph.D. in Computer Science from North Carolina State University. She conducts research in human factors in empirical software engineering including technical interviews, eye-movements in code review, and software developer productivity. Her current research investigates and devises intervention to support inclusive socio-technical interactions in online programming communities.
\end{IEEEbiography}
\vskip -1.5\baselineskip plus -1fil
\begin{IEEEbiography}[{\includegraphics[width=1in,height=2.5in,clip,keepaspectratio]{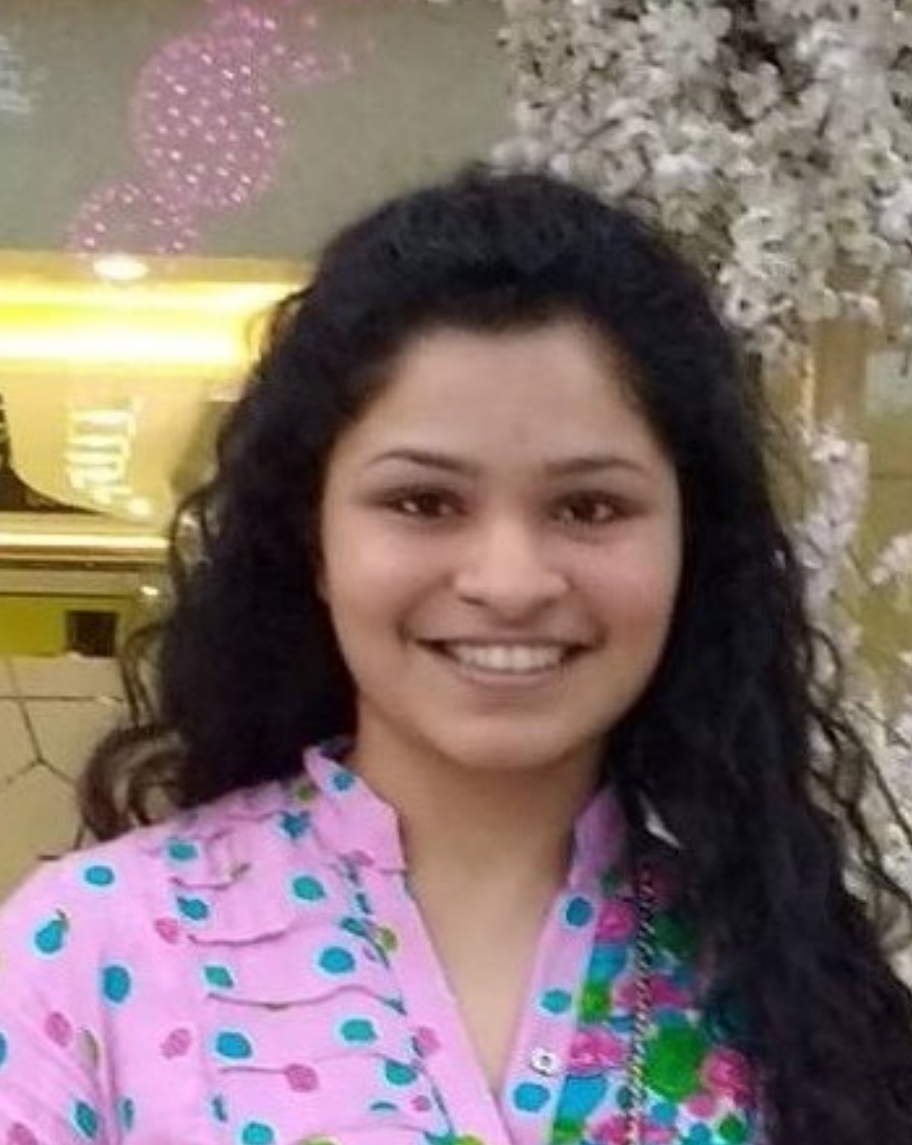}}]{Ayushi Rastogi}
is an Assistant Professor in the Faculty of Science and Engineering at the University of Groningen, the Netherlands. Her research interests include software analytics, empirical software engineering, and mining software repositories. She studies human and social aspects of software engineering for improving developer productivity and promoting diversity and inclusion. 
\end{IEEEbiography}
\vskip -2\baselineskip plus -1fil
\begin{IEEEbiography}[{\includegraphics[width=1in,height=1.25in,clip,keepaspectratio]{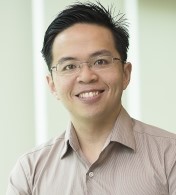}}]{David Lo}
is a Professor in the School of Computing and Information Systems at the Singapore Management University. He received his Ph.D. in Computer Science from the National University of Singapore. His research interests include software analytics, software maintenance, software testing, social network mining, and cybersecurity.
\end{IEEEbiography}
\vskip -2\baselineskip plus -1fil
\begin{IEEEbiography}[{\includegraphics[width=1in,height=1.25in,clip,keepaspectratio]{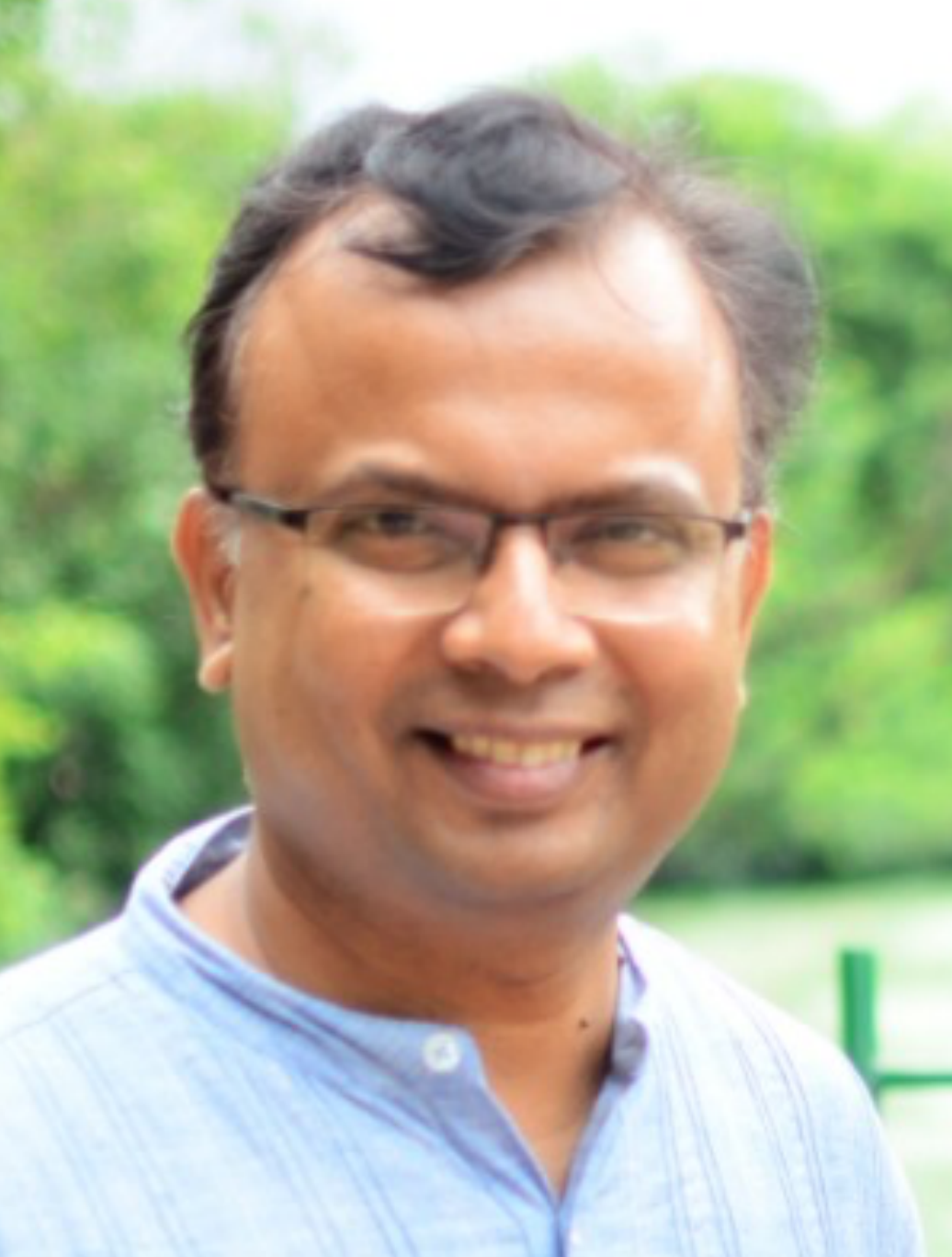}}]{Rahul Purandare}
is an Associate Professor in the department of Computer Science and Engineering at the Indraprastha Institute of Information Technology Delhi (IIIT-Delhi). He received his Ph.D. in Computer Science from the University of Nebraska - Lincoln. His research interests include program analysis, software testing, automatic program repair, code search, and code comprehension.
\end{IEEEbiography}
\vskip -1.5\baselineskip plus -1fil
\begin{IEEEbiography}[{\includegraphics[width=1in,height=1.5in,clip,keepaspectratio]{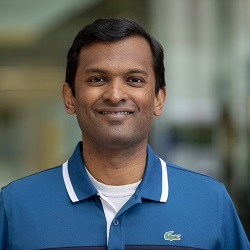}}]{Nachiappan Nagappan} was a Partner Researcher at Microsoft Research when this work was performed. He also holds an adjunct faculty appointment at IIIT New Delhi. His research interests are in the field of Data Analytics and Machine Learning for Software Engineering focusing on Analytics for Empirical Software Engineering, Software Reliability, Software Metrics, and Developer Productivity. He is a Fellow of the IEEE and ACM.
\end{IEEEbiography}

\end{document}